\documentstyle[12pt]{article}

\begin{document}

\title   {
                   Hexadecapole interaction and
                   the $\triangle I$=4 staggering effect
                   in rotational bands
         }

\author  {
                K. Burzy\'nski$^a$, P. Magierski$^b$, J. Dobaczewski$^a$,
                and W. Nazarewicz$^{a,b,c}$\\[3mm]
            $^a$Institute of Theoretical Physics,
                Warsaw University,\\
                Ho\.za 69, PL-00-681 Warsaw, Poland\\[3mm]
            $^b$Institute of Physics,
                Warsaw University of Technology,\\
                Koszykowa 75, PL-00-662 Warsaw, Poland\\[3mm]
            $^c$Joint Institute for Heavy Ion Research
                and Physics Division,\\
                Oak Ridge National Laboratory,
                Oak Ridge, Tennessee 37831, USA\\ and
                Department of Physics and Astronomy,\\ University of
                Tennessee, Knoxville, Tennessee 37996, USA
         }

\date{\it 17 October 1994}

\maketitle

\begin{abstract}
A role of the multipole interaction in  the description of the
$\triangle I$=4 staggering
phenomenon is investigated in a model consisting of a
single-$j$ shell filled by identical nucleons.
Exact diagonalization of the quadrupole-plus-hexadecapole Hamiltonian shows
that the hexadecapole-hexadecapole interaction can produce
a $\triangle I$=4 periodicity in the yrast sequence.
\end{abstract}

\section{Introduction}
\label{sec1}

Recently a $\triangle I$=4 staggering effect in the dynamical moment of
inertia of some superdeformed bands has been observed
\cite{Haa90,Fli93,Ced94}.
This can be interpreted as a bifurcation of the yrast band into two
sequences with spins varying by 4 units within each sequence.
The corresponding energy splitting is very small, i.e., about 100 eV.
This effect may occur due to remnants of the $C_4$ symmetry of the system
and suggests that presence of hexadecapole deformations or
hexadecapole-type multipole interactions may be responsible for the
staggering phenomenon. Recently, the  origin of the staggering
has been discussed by means of a phenomenological Hamiltonian
containing higher-order terms in angular momentum \cite{Ham94}.
In this work we investigate the effect by means of a more microscopic
approach involving schematic multipole-multipole interactions.

\section{Theoretical Model}
\label{sec2}

The model to test the $\triangle I$=4 staggering phenomenon should
be able to take into account the interplay between rotation
and shape dynamics of a many-body system.
A degenerate single-$j$ shell occupied by identical nucleons
 \cite{Bel59,MS64},
interacting
via quadrupole-quadrupole and hexadecapole-hexadecapole multipole forces
has this property.
The corresponding Hamiltonian may be written as
\begin{equation} \label{a1}
    \hat H = - \chi_2 \hat Q_2 \cdot \hat Q_2
             - \chi_4 \hat Q_4 \cdot \hat Q_4\; ,
\end{equation}
where $\hat Q_2$ and $\hat Q_4$ are the quadrupole and hexadecapole moment
operators defined as
$\hat Q_{\lambda\mu}$=$\sum_{mm'}(jmjm'|\lambda\mu) a_m^+ \tilde a_{m'}$
where $\tilde a_m$$\equiv$$Ta_mT^{-1}$, $T$ is the time reversal operator,
and the dot symbol  stands for the scalar product
  $\hat Q_\lambda$$\cdot$$\hat Q_\lambda$=$\sum_{\mu}(-)^{\mu}
  \hat{Q}_{\lambda\mu}\hat{Q}_{\lambda,-\mu}$.
In the harmonic oscillator approximation \cite{SK89}
the values of coupling constants $\chi_2$ and $\chi_4$
can be estimated as
\begin{equation}  \label{a15}
  \chi_{2}=\frac{176}{A} {\rm~MeV}\; , \qquad\qquad
  \chi_{4}=\frac{165}{A} {\rm~MeV}\; .
\end{equation}
The results of the diagonalization depend solely on the ratio
$\chi_4/\chi_2$.
Therefore, in the following, we assumed $\chi_2$=1~MeV
(i.e., $\chi_2$ is used to fix energy scale),
while the parameter
 $\chi_4$,
describing the  relative strength of hexadecapole and quadrupole
interactions,
was varied to investigate the conditions for
an appearance
of the staggering effect.

In this paper we discuss the case of $N$=8 particles in the $j$=15/2 shell,
although we have performed a series of calculations for several
different shells and particle numbers.
A choice of $N$=8 particles corresponds to a half-filled shell,
and gives a regular collective yrast band.
The calculations were performed for the parameter $\chi_4$
ranging from $\chi_4$=0 to $\chi_4$=10~MeV.
For larger values of $\chi_4$, one only obtains an energy scaling
corresponding to the dominating hexadecapole interaction.

To analyse a $\triangle I$=4 staggering effect in collective bands,
we extracted the smooth reference curve according to Ref.{\ }\cite{Fli93}.
Quantities
$\triangle E_\gamma(I)$$\equiv$$E_\gamma(I$$+$$2)$$-$$E_\gamma(I)$
and
$\triangle E_\gamma^{\rm ref}(I)$$\equiv$
$[\triangle E_\gamma(I$$+$$2)$$+$$2
 \triangle E_\gamma(I)$$+$$\triangle E_\gamma(I$$-$$2)]/4$
were obtained in this way.
Then staggering parameter
$\triangle E_\gamma (I)$$-$$\triangle E_\gamma^{\rm ref}(I)$
was evaluated and plotted in each case.

\section{Results and Discussion}
\label{sec3}

Fig.{\ }\ref{f1} shows a complete quadrupole spectrum of Hamiltonian
(\ref{a1}) in case of $\chi_4$=0.
Solid lines connect states of spins $I$ and $I$$+$2 with the
largest reduced matrix
elements of the quadrupole moment operator.
Several quite regular rotational bands were obtained.
The staggering parameter was evaluated for each case,
but no $\triangle I$=4 periodicity was found.
In Figs.{\ }\ref{f2-3}a and \ref{f2-3}b the collective energies $E(I)$
and the parameters
$\triangle E_\gamma (I)$$-$$\triangle E_\gamma^{\rm ref}(I)$
of the yrast sequence are presented as open circles.
When the hexadecapole interaction is switched on, the staggering
$\triangle I$=4 appears.
This is seen in Fig.{\ }\ref{f2-3}b, where full circles denote
the staggering parameters of the yrast band for $\chi_4$=0.3, 0.4,
and 0.5~MeV,
while the corresponding yrast spectra are plotted in Fig.{\ }\ref{f2-3}a.
The staggering effect is small and therefore difficult to be directly seen
in the yrast spectra.

The amplitude of the staggering increases with $\chi_4$,
but for the values of $\chi_4$ larger than 0.5~MeV,
the yrast states are no longer connected by the enhanced E2 transitions;
i.e., the band structure is lost.
Fig.{\ }\ref{f4-5}a shows the yrast states for $\chi_4$=4~MeV.
This corresponds to the situation where the hexadecapole interaction
dominates and the yrast sequence is strongly perturbed.
There are no strong E2 transitions linking the yrast
states, but the staggering
parameter extracted from the energies of the yrast sequence exhibits
the $\triangle I$=4 periodicity (Fig.{\ }\ref{f4-5}b).

Similar effects can be obtained for other values of $j$ and $N$.
We would like to stress that it was impossible to obtain similar staggering
by using the quadrupole-quadrupole interaction alone.
On the other hand, for $j$=15/2 and $N$=8,
the staggering appears only in the yrast band.

The examples presented here show
that the hexadecapole interaction may generate the $\triangle I$=4
irregularities.
The proper treatment of the hexadecapole interaction
seems to be crucial for the understanding of this
intriguing  staggering phenomenon.

\section*{Acknowledgements}

This research was supported in part by the
Polish State Committee for Scientific Research under Contract
Nos. 20450~91~01 and 2~P~302~056~06,
and by the computational grant from
The Interdisciplinary Centre for Mathematical and Computational Modeling
(ICM) of Warsaw University.
Oak Ridge National
Laboratory is managed for the U.S. Department of Energy by Martin
 Marietta Energy Systems, Inc. under Contract No.
DE-AC05--84OR21400.
The Joint Institute for Heavy Ion
 Research has as member institutions the University of Tennessee,
Vanderbilt University, and the Oak Ridge National Laboratory; it
is supported by the members and by the Department of Energy
through Contract No. DE-FG05-87ER40361 with the University
of Tennessee.  Theoretical nuclear physics research
at the University of Tennessee
 is supported by the U.S. Department of
Energy through Contract No. DE-FG05-93ER40770.

\newpage
\begin{figure}[h]
\caption[9]{The exact spectrum for the $j$=15/2
            shell filled with $N$=8 particles.  The coupling constants
            $\chi_2$=1~MeV, $\chi_4$=0 were used.}
\label{f1}
\end{figure}

\begin{figure}[h]
\caption[9]{The results for the $j$=15/2 shell filled with $N$=8
            particles.  The quadrupole coupling constant was $\chi_2$=1~MeV.
            The hexadecapole strengths $\chi_4$=0 (open circles) and
            $\chi_4$=0.3, 0.4, 0.5~MeV (full circles) were used.
            Part (a) shows the yrast spectra, part (b) the corresponding
            staggering parameters
            $\triangle E_\gamma(I)$$-$$\triangle E_\gamma^{\rm ref}(I)$
            plotted as a function of spin.}
\label{f2-3}
\end{figure}

\begin{figure}[h]
\caption[9]{The exact yrast spectrum (a) and the corresponding
            staggering parameter
            $\triangle E_\gamma(I)$$-$$\triangle E_\gamma^{\rm ref}(I)$ (b)
            for the $j$=15/2 shell filled with $N$=8 particles.
            The quadrupole and hexadecapole coupling constants
            were $\chi_2$=1~MeV, $\chi_4$=4~MeV, respectively.}
\label{f4-5}
\end{figure}

\end{document}